\documentclass[11pt]{article}
\usepackage{SIGACTNews}
\usepackage{listings}
\usepackage{graphicx}
\usepackage{amssymb}
\usepackage[boxed]{algorithm}
\usepackage{algorithmic}
\usepackage{amsmath}
\newtheorem{thm}{Theorem}[section]

\newcounter{theoremcounter}[section] 
\renewcommand{\thetheoremcounter}{\thesection.\arabic{theoremcounter}}

\newenvironment{theoremstatement}%
               {\vspace{0.01\textwidth}\par\noindent\par\nopagebreak\par\nopagebreak
               \setlength{\parindent}{0em}%
               \setlength{\parskip}{1ex}%
               \refstepcounter{theoremcounter}%
               \textbf{Theorem \nopagebreak[2]\thetheoremcounter{}: }}{\vspace{0.1in}\par}

               {\par\noindent%
               \setlength{\parindent}{0em}%
               \setlength{\parskip}{1ex}%
               \textbf{Proof:\nopagebreak[2]}}{\hfill\rule{1.5ex}{1.5ex}\vspace{0.1in}\par}

\newcounter{lemmacounter}[section] 
\renewcommand{\thelemmacounter}{\thesection.\arabic{lemmacounter}}

\newenvironment{lemmastatement}%
               {\vspace{0.01in}\par\noindent\par\nopagebreak\par\nopagebreak
               \setlength{\parindent}{0em}%
               \setlength{\parskip}{1ex}%
               \refstepcounter{lemmacounter}%
               \textbf{Lemma \nopagebreak[2]\thelemmacounter{}: }}{\vspace{0.1in}\par} 
\newenvironment{lemmaproof}%
               {\par\noindent%
               \setlength{\parindent}{0em}%
               \setlength{\parskip}{1ex}%
               \textbf{Proof:\nopagebreak[2]}}{\hfill\rule{1.5ex}{1.5ex}\vspace{0.1in}\par}

\newtheorem{obs}{Observation}[section]

\newtheorem{def1}[thm]{Definition}

\newcommand{\BibTeX}{{\rm B\kern-.05em{\sc i\kern-.025em b}\kern-.08em
    T\kern-.1667em\lower.7ex\hbox{E}\kern-.125emX}}

\newenvironment{deflist2}{
\begin{itemize}
                      \setlength{\itemsep}{0pt}
                      \setlength{\parskip}{0pt}
                      \setlength{\parsep}{0pt}
                      \setlength{\itemindent}{1pt}}{\end{itemize}
}

\title{Improved Fully Dynamic Reachability Algorithm for Directed Graph}
\author{Venkata Seshu Kumar Kurapati \footnote{The views expressed in this paper are the author's own and do not necessarily represent the views of Microsoft} \\ \small{seshuk@microsoft.com} }
\date{03 Feb 2005}
\begin{document}
\SIGACTmaketitle
\setcounter{page}{1}
\begin{abstract}
We propose a fully dynamic algorithm for maintaining reachability information in directed graphs. The proposed deterministic dynamic algorithm has an update time of $O( (ins*n^{2}) + (del * (m+n*log(n))) )$ where $m$ is the current number of edges, $n$ is the number of vertices in the graph, $ins$ is the number of edge insertions and $del$ is the number of edge deletions. Each query can be answered in $O(1)$ time after each update. The proposed algorithm combines existing fully dynamic reachability algorithm with well known witness counting technique to improve efficiency of maintaining reachability information when edges are deleted. The proposed algorithm improves by a factor of $O(\frac{n^2}{m+n*log(n)})$ for edge deletion over the best existing fully dynamic algorithm for maintaining reachability information.
\end{abstract}
\setcounter{page}{1}
\section{Introduction}
Transitive Closure problem for directed graph is well studied and motivated problem. Many applications either can be represented using directed graphs or use directed graph as underlying data structure. In reality for many of these applications, underlying directed graph changes dynamically i.e. edges are inserted and/or deleted. This is a typical scenario in network routing, transportation and data mining. This leads to answering reachability queries for changing graph. As with any dynamic algorithm, we have two values in mind to understand efficiency of dynamic algorithm i.e. update time and query time. Update time is amount of effort required to handle changes to graph while maintaining some information(reachability) related to the graph and query time is amount of effort required to answer (reachability) query on current graph after each update. Dynamic algorithms for maintaining reachability information on directed graphs has taken two paths in recent past. The trade off between these two approaches lies in update time vs. query time. First set of algorithms maintain reachability information explicitly in terms of transitive closure matrix where as the second set of algorithms don't explicitly maintain reachability information in the form of transitive closure matrix.  First set of algorithms will have a query time of $O(1)$ as each query is a look up operation on the transitive closure matrix which is $O(1)$ after each update.
\par
Assuming reachability information is maintained explicitly in a transitive closure matrix then insertion or deletion of an edge operation may change $O(n^2)$ entries in the transitive closure matrix. Therefore an amortized update time of $O(n^2)$ is essentially optimal for insert or delete operation. This optimal result is achieved by Roditty\cite{L03}, Demetrescu and Italiano\cite{DI00} as their algorithms maintain transitive closure of directed graph with an amortized update time of $O(n^2)$ by improving the algorithm proposed by King\cite{K99}. In both these optimal results, each query can be answered in time $O(1)$ as transitive closure is explicitly maintained for the directed graph in consideration. Also both these algorithms support extended insert(may insert a set of edges all touching one vertex) operation and fully extended delete operation where completely arbitrary set of edges may be deleted from the graph in one delete operation.
\par
The second set of algorithms, where transitive closure matrix is not explicitly maintained, motivated by the fact that if number of queries after each update operation is relatively small, then we can have a dynamic algorithm with a smaller update time  at the expense of a larger query time for maintaining reachability information. Roditty and Zwick\cite{RZ02} proposed a fully dynamic algorithm of this type where update time of the algorithm is $O(m\sqrt{n})$ and answers queries in $O(\sqrt{n})$ time. They achieve this by improving the results of Henzinger and King\cite{HK95}. Demetrescu and Italiano\cite{DI00} obtained a dynamic algorithm based on fast matrix multiplication whose amortized update time is $O(n^{1.575})$ and query time is $O(n^{0.575})$. Their algorithm\cite{DI00} is a randomized algorithm having one sided error and only works for directed acyclic graphs(DAGs).
\par
Roditty and Zwick\cite{RZ04} obtained an algorithm for maintaining reachability information with an amortized almost linear update time with respect to the number of edges and vertices in the current graph i.e. $O(m+n*log(n))$ while it can answer queries in $O(n)$ time after each update. They obtained this algorithmic result by proposing an efficient way of maintaining strongly connected components of a directed graph which supports an interesting persistence property. Their algorithm is the first algorithm that breaks the $O(n^2)$ update barrier for all graphs with $o(n^2)$ edges with query time equal to $O(n)$. 
\par
The proposed algorithm has an update time of $O( (ins*n^{2}) + (del * (m+n*log(n))) )$ where $m$ is the current number of edges, $n$ is the number of vertices in the graph, $ins$ is the number of edge insertions and $del$ is the number of edge deletions. Each query can be answered in $O(1)$ time after each update. Proposed algorithm combines algorithm presented in \cite{RZ04} with well known witness counting technique to improve update time for edge deletion form $O(n^2)$ to $O(m+n*log(n)))$ while answering queries in $O(1)$ time instead of $O(n)$.
\par
The proposed algorithm and some of the existing algorithms for dynamic reachability problem are compared in Table~\ref{t1}.
\begin{table}[h]
\centering
\footnotesize
\label{t1}
\begin{tabular}{||r|r|r|r|r|r|}
\hline
Type of Graph & Type of Algorithm & Insertion Time & Deletion Time & Query Time & Reference \\
 & & (Amortized) &  (Amortized) & & \\
\hline
\hline
General & Deterministic & $O(n^2)$ & $O(n^2)$ & $O(1)$ & \cite{DI00}\cite{L03} \\
General & Monte Carlo & $O(m*\sqrt{n}*log^2(n))$ & $O(m*\sqrt{n}*log^2(n))$& $O(n/log(n))$ & \cite{HK95} \\
General & Deterministic & $O(m*\sqrt{n})$ & $O(m*\sqrt{n})$ & $O(\sqrt{n})$ & \cite{RZ02} \\
General &  Monte Carlo & $O(m^{0.58}*n)$ & $O(m^{0.58}*n)$ & $O(m^{0.43})$ & \cite{RZ02} \\
DAGS    &  Monte Carlo & $O(n^{1.575})$ & $O(n^{1.575})$ & $O(n^{0.575})$ & \cite{DI00} \\
General &  Monte Carlo & $O(n^{1.575})$ & $O(n^{1.575})$ & $O(n^{0.575})$ & \cite{PIO04} \\
General & Deterministic & $O(m+n*log(n))$ & $O(m+n*log(n))$ & $O(n)$ & \cite{RZ04} \\
\hline
General & Deterministic & $O(n^2)$ & $O(m+n*log(n))$ & $O(1)$ & In this paper \\
\hline
\end{tabular}
\normalsize
\caption{Fully Dynamic Reachability Algorithms}
\end{table}
Rest of this paper is organized as follows. In the next section we introduce required concepts which are going to be used in the proposed algorithm and review some of algorithms which are going to be refined in later sections. In Section 3 we lazily update transitive closure using well known witness counting technique. Section 4 analyzes time complexity of proposed algorithm. We end this paper by concluding remarks and open problems in Section 5.
\section{Background}
Let $G(V,E)$ be a directed graph where $V$ is set of nodes of the graph and $E$ is set of edges of the graph where $(v,u) \in E$ iff $v,u \in V$ and there is a direct edge between nodes $v,u$ in $G$. Each node in the graph is given unique value whose range is between $1,|V|$. We most of the time use this value as the node itself when considering data structures. Most of other definitions and notations are borrowed from Roditty and Zwick\cite{RZ04}. 
\subsection{Graph Sequence}
\begin{figure}[h]
\centering
\framebox{\begin{minipage}{0.65\textwidth}
   \begin{deflist2}
	  	\item 
	  		Insert($E^{'}$): $t \leftarrow t + 1$,$E_t \leftarrow E_{t-1} \cup E^{'} $
	  	\item
	  		Delete($E^{'}$): $E_j \leftarrow E_{j} - E^{'}$, for 	$1 \leq j \leq t$ 
	  \end{deflist2}
 \end{minipage}}
 \caption{Insert and Delete operations on graph sequence for graph $G(V,E)$}
 \label{ldef}
\end{figure}
\par
Since we are dealing with dynamic problem i.e. maintaining reachability information when $G(V,E)$ under goes edge insertion and deletions, we define the following sequence on the graph $G(V,E)$. Let $G_1,G_2,...,G_t$ denote sequence of graph versions of $G(V,E)$ which has undergone edge insertions and deletions where $t$ is the number of insertions performed so far. $G_i(V,E_i)$ is created by $i^{th}$ insert operation where $0 \leq i \leq t$. Let us assume $G_0(V,E_0)$ as empty graph i.e. $E_0={\phi}$ with out loss of generality. The creation of a new version in the graph sequence only happens due to edge insertion.
\par
We will never create a new version of the graph for a delete operation. During delete operations, we will update all the versions of the graph by deleting edges if they are present in that particular version of the graph i.e. if we want to delete edge $(v,u)$ and if $(v,u) \in G_i$ where $1 \leq i \leq t$, then we delete edge $(v,u)$ from $G_i(V,E_{i})$. 
\par
Proposed algorithms can support extended insert and extended delete operations where each extended operation is rooted at a particular node $v$ of the graph $G(V,E)$.  We introduce the notion of time line for the graph sequence in order to explain algorithm and concepts in an efficient manner. Each update operation on the graph increases time line of the graph by one. This time line is denoted by $Tline_G$. Note that an insertion center $G(V,E_i)$ may be inserted during time line $k$ where $k \geq i$. This time line is not used in proposed algorithms but used to simplify proofs and explanation.
\subsection{In/Out Trees and Decremental Maintenance of In/Out Trees}
The proposed fully dynamic algorithms work on the ability of maintaining reachability trees for each version of the graph rooted at an extended insert operation root. 
\begin{def1}
A reachability tree is used to maintain the set of vertices that are reachable from a certain vertex $r$ of a graph $G = (V,E)$
that undergoes a sequence of edge deletions\cite{RZ04}. The nodes of the tree are not individual vertices of $G$ but rather strongly connected components of $G$. The root of the tree is the component containing $r$.
\end{def1} 
$In[u]$ denotes the reachability in tree rooted at node $u$ i.e. $In[u]$ is a tree containing all the nodes in $G(V,E)$ which can reach node $u$. We similarly denote $Out[u]$ as the reachability out tree rooted at node $u$ i.e. $Out[u]$ is a tree containing all the nodes of $G(V,E)$ which are reachable from node $u$. From\cite{RZ04}, we have the following theorem which is one of the ingredients of the proposed algorithm. We say $ReachIn(u,r)$ is true if $u \in In[r]$ and similarly $ReachOut(u,r)$ is true if $u \in Out[r]$ otherwise false.
\begin{theoremstatement}
The total time needed to maintain a reachability in(out)tree rooted at $r$ is only $O(m + n*log(n))$ when the graph $G(V,E)$ under goes a sequence of edge deletions whereas the queries $ReachIn(u,r)$($ReachOut(u,r)$) can be answered in $O(1)$ time after each edge deletion\cite{RZ04}.
\end{theoremstatement}
In order to answer reachability query on the current version of the graph $G(V,E_t)$, we use the following approach. $Query(v,u)$ is true iff $\exists{r_i}$ such that $v \in In[r_i]$ and $u \in Out[r_i]$ where $r_i$ denotes insertion center for $i^{th}$ insertion operation and $0 \leq i \leq t$. Otherwise $Query(v,u)$ is false. We can answer reachability queries this way on current version of the graph $G(V,E_t)$. Since there can be at most $O(n)$ insertion centers for any arbitrary sequence of edge insertions and edge deletions, we can have $O(n)$ as query time after each update. In the next section we show that using the concept of lazy update how to reduce query time after each update.
\section{Counting Technique}
For time being, let $G(V,E)$ be any graph that is undergoing a sequence of edge deletions only. We say a node $v$ is part of $In[r]$ iff $ReachIn(v,r)$ is true in the current version of the graph. Similarly we say a node $v$ is part of $Out[r]$ iff $ReachOut(v,r)$ is true in the current version of the graph and is false otherwise.
\begin{obs}
Let $r_i$ be an insertion center during time line $k$ where $k \geq i$ and $r_i$ becomes again an insertion center at time line $l (> k)$. Let us say a node $v \in In[r_i]$ at time line $k$ then $v$ will be deleted only once from $In[r_i]$ before time line $l$. This happens only once when the node $v$ is not reachable from node $r_i$ where $r_i$ is $i^{th}$ insertion center and $0 \leq i \leq t$. Similarly this is applicable to $Out$ arrays.
\label{l1}
\end{obs}
We use above observation~\ref{l1} and maintain transitive closure matrix $TCM$.
\begin{table}[h]
\centering
\setlength{\abovecaptionskip}{0pt} 
\setlength{\belowcaptionskip}{0pt} 
 \framebox{\begin{minipage}{0.6\textwidth}
	  \begin{algorithmic}[1] 
	  	\STATE {\bf Input:} $v,u$
	  	\IF{ $TCM[v][u]>0$}
	  		\STATE return true
	  	\ELSE
	  		\STATE return false
	  	\ENDIF
	  \end{algorithmic}
	  \caption{$Query(v,u)$}
	  \label{algoq1}
 \end{minipage}}
\end{table}
Using $TCM$, we can answer $Query(v,u)$ based on $TCM$ and is shown in Algorithm~\ref{algoq1}. $Query(v,u) = 1$ if $TCM[v][u]>0$ otherwise $0$. Note that $TCM$ is not a binary matrix but contains number of insertion centers who can witness for a path directed from node $v$ to node $u$ when we are looking at $TCM[v][u]$. Note that there can not be a node $z$, which is not an insertion center, can witness a directed path from a node to another node as there will not be outgoing edges from node $z$. If not, then $z$ must be a insertion center at some point along time line. In the next few paragraphs, we show how to maintain witness count after each update. We refer to term "witness count" as number of insertion centers witnessing a directed path between two nodes in consideration.
\subsection{Handling Insertions}
\begin{table}[h]
\centering
\setlength{\abovecaptionskip}{0pt} 
\setlength{\belowcaptionskip}{0pt} 
 \framebox{\begin{minipage}{0.8\textwidth}
	  \begin{algorithmic}[1] 
	  	\STATE {\bf Input:} $E_{v}$
	  	\IF{ $v$ is a previous insertion center}
	  		\FOR{each $u$ in $In[v]$}
		  		\FOR{each $z$ in $Out[v]$}
		  			\STATE $TCM [u][z] = TCM [u][z] - 1$
		  		\ENDFOR
	  	\ENDFOR
	  	\ENDIF
	  	\STATE compute new $In$ and $Out$ trees rooted at node $v$ using algorithm mentioned in \cite{RZ04}
	  	\FOR{each $u$ in $In[v]$}
	  		\FOR{each $z$ in $Out[v]$}
	  			\STATE $TCM [u][z] = TCM [u][z] + 1$
	  		\ENDFOR
	  	\ENDFOR
	  \end{algorithmic}
	  \caption{$Insert(E_{v})$}
	  \label{algoq2}
 \end{minipage}}
 \end{table}
We handle insertion as follows. Let us say we want to insert a set of edges, $E_v$, centered at node $v$. Let us assume $v$ is not a insertion center before. First step of algorithm is to compute $In[v]$ and $Out[v]$ for current graph using algorithm mentioned in \cite{RZ04} so that both $In[v]$ and $Out[v]$ can be maintained when edges are deleted from graph $G(V,E)$ in $O(m+n*log(n))$ time where $m$ is current number of edges in the graph and $n$ is number of vertices in the graph. Using $In[v]$ and $Out[v]$, we update witness count as follows. For each node $u \in In[v]$ and $z \in Out[v]$, we know that node $v$ witness a path from node $u$ to node $z$. So we update witness count for a directed path from node $u$ to node $z$ by increasing $TCM[u][z]$ by one. 
\par
Now we remove assumption that $v$ is not a insertion center before. To handle this case, just before computing $In[v]$ and $Out[v]$, we remove node $v$ as a witness as we are going to compute from scratch when node $v$ acts as an witness. So we decrement witness count by one for each pair $(v,u)$ where $v \in In_[v]$ and $u \in Out[v]$. This we perform before we compute new $In[v]$ and $Out[v]$. Algorithm shown in Algorithm~\ref{algoq2} shows how to update $TCM$ after inserting a set of edges, $E_v$, centered at node $v$.
\begin{lemmastatement}
Algorithm shown in Algorithm~\ref{algoq2} correctly updates $TCM$ where $TCM[u][z]$ contains number of insertion centers witnessing a directed path from node $u$ to node $z$ after inserting a set of edges, $E_v$, centered at node $v$.
\end{lemmastatement}
\begin{lemmaproof}
We have two cases.
\begin{enumerate}
\item 
Let us assume $v$ is not an insertion center before. We know for each $u \in In[v]$ and $z \in Out[v]$, there is a directed path from node $u$ to node $z$ via node $v$. So node $v$ becomes a witnessing insertion center for a directed path from node $u$ to node $z$. This is exactly what is performed by algorithm shown in Algorithm~\ref{algoq2} by incrementing $TCM[u][z]$ with one.
\item
Let us say $v$ is an insertion center before. Let $In_{old}[v]$ and $Out_{old}[v]$ denote In and Out trees rooted at node $v$ just before inserting edge set $E_v$. $In_{new}[v]$ and $Out_{new}[v]$ denote In and Out trees rooted at node $v$ after inserting edge set $E_v$. We know that we need to update $TCM[u][z]$ if only if 
\begin{enumerate}
\item $u \in In_{new}[v] - In_{old}[v]$ and $z \in Out_{new}[v]$
\item $u \in In_{old}[v]$ and $z \in Out_{new}[v] - Out_{old}[v]$
\end{enumerate}
This is true as inserting edge set $E_{v}$, can only create additional paths between nodes which were not present in $In_{old}[v]$ and present in $In_{new}[v]$ to any node in $Out_{new}[v]$. Similarly we can have new path between nodes which were present in $In_{old}[v]$ and to nodes which are presented in $Out_{new}[v]$ but not in $Out_{old}[v]$. Note that with out these new paths, node $v$ can not act as insertion center witnessing between above stated combination of nodes.
\par
This is exactly performed in algorithm shown in Algorithm~\ref{algoq2} but in a different way by first decrementing $TCM[u][z]$ for each $u \in In_{old}[v]$ and $z \in Out_{old}[v]$ and then incrementing $TCM[u][z]$ for each $u \in In_{new}[v]$ and $z \in Out_{new}[v]$ but both operations are equal.
\end{enumerate}
\end{lemmaproof}
\subsection{Handling Deletions}
\begin{table}[h]
\centering
\setlength{\abovecaptionskip}{0pt} 
\setlength{\belowcaptionskip}{0pt} 
 \framebox{\begin{minipage}{\textwidth}
	  \begin{algorithmic}[1] 
	  	\STATE {\bf Input:} $E_{v}$
	  	\FOR{each version of graph $G_i$ rooted at $z$}
	  		\STATE update $In[z]$ and $Out[z]$ using algorithm mentioned in \cite{RZ04}
	  		\STATE denote $In_{delete}[z]$ as nodes which are deleted from $In[z]$ due to deletion of edge set $E_{v}$.
	  		\STATE denote $Out_{delete}[z]$ as nodes which are deleted from $Out[z]$ due to deletion of edge set $E_{v}$.
	  	\ENDFOR
	  	\STATE Call UpdateTCM
	  \end{algorithmic}
	  \caption{$Delete(E_{v})$}
	  \label{algoq3}
 \end{minipage}}
\end{table} 
We handle edge deletions as follows.  Let us say we want to delete a set of edges, $E_v$, centered at node $v$. We know that node $v$ is an insertion center. Let us assume that $In[v]$ is set of nodes which can reach node $v$ and $Out[v]$ is set of nodes which can be reachable from node $v$ just before deleting edge set $E_v$. Let $In_{update}[v]$ denotes set of nodes which can reachable from node $v$ and $Out_{update}[v]$ is set of nodes which can be reachable from node $v$ after deletion of edge set $E_v$. 
\[ In_{delete}[v] = In[v] - In_{update}[v] \]
\[ Out_{delete}[v] = Out[v] - Out_{update}[v] \]
$In_{delete}[v]$ and $Out_{delete}[v]$ denotes set of nodes whose reachability got effected by deletion of edge set $E_{v}$ with respect to node $v$. We use following observations to handle edge deletions.
\begin{obs}
Let us assume $u \in In[v]$. After deleting edge set $E_v$, node $v$ will not act as a witness for directed path from node $u$ to node $z$ any more iff $u \in In_{delete}[v]$ and $z \in Out[v]$. 
\end{obs}
\begin{obs}
Let us assume $z \in Out[v]$. After deleting edge set $E_v$, node $v$ will not act as a witness for directed path from node $u$ to node $z$ any more iff $u \in In[v]$ and $z \in Out_{delete}[v]$. 
\end{obs}
Given $In_{delete}[v]$ and $Out_{delete}[v]$, we can update $TCM$ using above two observations. Algorithm shown in Algorithm~\ref{algoq3} shows how to update $TCM$ after deleting a set of edges, $E_v$, centered at node $v$. First step of algorithm is to compute $In_{delete}[v]$ and $Out_{delete}[v]$ which can be easily obtained by simple modification to algorithm presented in \cite{RZ04} for maintaining In and Out trees rooted at a node $v$ when graph is undergoing edge deletions. We use above two observations and update $TCM$ by just one iteration over $In_{delete}[v]$, $In[v]$, $Out[v]$ and $Out_{delete}[v]$. This is handled in algorithm $UpdateTCM$ shown in Algorithm~\ref{algoq4}.
\begin{table}[h]
\centering
\setlength{\abovecaptionskip}{0pt} 
\setlength{\belowcaptionskip}{0pt} 
 \framebox{\begin{minipage}{0.6\textwidth}
	  \begin{algorithmic}[1] 
	  	\FOR{each version of graph $G_i$ rooted at $z$}
		  	\FOR{each $u$ in $In_{delete}[z]$}
			  		\FOR{each $l$ in $Out[z]$}
			  			\STATE $TCM [u][l] = TCM [u][l] - 1$
			  		\ENDFOR
		  	\ENDFOR
		  \ENDFOR
		  \FOR{each version of graph $G_i$ rooted at $z$}
		  	\FOR{each $u$ in $Out_{delete}[z]$}
			  		\FOR{each $l$ in $In[z]$}
			  			\STATE $TCM [l][u] = TCM [l][u] - 1$
			  		\ENDFOR
		  	\ENDFOR
		  \ENDFOR
	  \end{algorithmic}
	  \caption{$UpdateTCM$}
	  \label{algoq4}
 \end{minipage}}
\end{table} 
\begin{lemmastatement}
Algorithm shown in Algorithm~\ref{algoq3} correctly updates $TCM$ where $TCM[u][z]$ contains number of insertion centers witnessing a directed path from node $u$ to node $z$ after deleting a set of edges, $E_v$, centered at node $v$ given $In_{delete}[z]$ and $Out_{delete}[z]$ for each insertion center $z$.
\end{lemmastatement}
\begin{lemmaproof}
Let $l$ be any insertion center and is witnessing a directed path from node $u$ to node $z$ before deletion of edge set $E_{v}$. This witness will cease if $u \in In_{delete}[l]$ and $z \in Out[l]$ as $u$ can not reach node $l$ after deletion of edge set $E_v$. This is even true in case $u \in In[l]$ and $z \in Out_{delete}[l]$ as $z$ can not be reached from node $l$ after deletion of edge set $E_v$. $l$ will remain as witness when $u \in In_{update}[l]$ and $z \in Out_{update}[l]$. This is exactly performed in algorithm shown in Algorithm~\ref{algoq3}.
\end{lemmaproof}
We combine algorithm~\ref{algoq1},~\ref{algoq2} and algorithm~\ref{algoq3} to obtain a fully dynamic algorithm for maintaining transitive closure. We can obtain $In_{delete}[z]$ and $Out_{delete}[z]$ for each insertion center after edge deletion by trivial modification to algorithm presented in \cite{RZ04} for decremental maintenance of In and Out trees rooted at node $z$.
\section{Time Complexity}
In this section, we analyze time complexity of algorithm for maintaining transitive closure presented in previous section when graph is under going edge insertions and deletions.
\begin{lemmastatement}
Time complexity of Algorithm shown in Algorithm~\ref{algoq3} is $O(n^2)$.
\end{lemmastatement}
\begin{lemmaproof}
There can be at most $n$ nodes in $In[v]$ and at most $n$ nodes in $Out[v]$. Therefore there can be at most $n^2$ updates to $TCM$ where $v$ is not an insertion center before. If $v$ is an insertion center before, then at most $n^2$ entries will be decremented in $TCM$ and at most $n^2$ entries will be incremented for new insertion center rooted at node $v$. Therefore proves theorem.
\end{lemmaproof}
\begin{lemmastatement}
For an insertion center $v$, number of entries changed in $TCM$ while maintaining In and Out trees rooted at node $v$ is $O(n^2)$.
\end{lemmastatement}
\begin{lemmaproof}
There can be at most $n$ nodes in In tree and at most $n$ nodes in Out tree rooted at node $v$ when node $v$ becomes an insertion center. After node $v$ became an insertion center and just before it again becomes an insertion center along time line, nodes will disappear from In and Out tree rooted at node $v$ as there can be only deletion of edges from In and Out tree rooted at node $v$. In worst case, there will not be a singe node in both In and Out tree just before node $v$ again becomes an insertion center. Therefore node $v$ will cease to witness at most $n^2$ pairs directed path via node $v$. Therefore proves theorem.
\end{lemmaproof}
\begin{lemmastatement}
Let $v$ be any insertion center, then computing $In_{delete}[v]$ and $Out_{delete}[v]$ while maintaining In and Out trees rooted at node $v$ takes at most $O(m+n*log(n))$ time using decremental maintenance of In and Out trees algorithm mentioned in \cite{RZ04}
\end{lemmastatement}
\begin{lemmaproof}
From \cite{RZ04}, we know that In and Out trees can be maintained decrementally rooted at an insertion center in $O(m+n*log(n))$ time. We can easily obtain $In_{delete}[v]$ and $Out_{delete}[v]$ after each edge deletion by simply denoting nodes explicitly which are not present in both In and Out trees after each edge set deletion but are present before edge set deletion.
\end{lemmaproof}
\begin{lemmastatement}
Combining algorithm~\ref{algoq1},~\ref{algoq2} and algorithm~\ref{algoq3}, we obtain a fully dynamic algorithm for maintaining transitive closure with $O(1)$ query time where insertion update time and deletion update time are $O(n^2)$ and $O(m+n*log(n))$ respectively.
\end{lemmastatement}
\begin{lemmaproof}
From lemma 4.1,we know that time taken to maintain $TCM$ when we insert a set of edges rooted at node $v$ is $O(n^2)$ and also decremental maintenance of $In[v]$ and $Out[v]$ trees rooted at insertion center $v$ is $O(m+n*log(n))$ which includes cost for providing $In_{delete}[v]$ and $Out_{delete}[v]$ after each update. We know from lemma 4.2, there can be at most $O(n^2)$ updates to $TCM$ for an insertion center $v$ where $In[v]$ and $Out[v]$ are undergoing edge deletions only. This cost is amortized onto insertion. Therefore proves theorem.
\end{lemmaproof}
\section{Future work and Conclusions}
In this paper we have proposed an algorithm for maintaining reachability information when graph is undergoing edge insertions and deletions. The algorithm has an insertion update time of $O(n^2)$ and a deletion update time of $O(m+n*log(n))$. Queries can be answered in $O(1)$ time after each update. This is the first deterministic algorithm to have linear update time for edge deletion while answering queries in $O(1)$ time. The proposed algorithm improves by a factor of $O(\frac{n^2}{m+n*log(n)})$ over existing best fully dynamic algorithm for maintaining reachability information. Reducing insertion time while maintaining same deletion update time and query time is a major open problem.
\bibliographystyle{alpha} 
\bibliography{bib}
\end{document}